\documentclass{revtex4}
\usepackage{graphicx}
\usepackage{amsmath}
\usepackage{amssymb}

\def\Z{{\mathbb Z}}
\begin{document}
\title{4D static solutions with interacting phantom fields}
\author{Vladimir Dzhunushaliev
\footnote{Senior Associate of the Abdus Salam ICTP}}
\email{dzhun@krsu.edu.kg}
\affiliation{Dept. Phys. and Microel.
Engineer., Kyrgyz-Russian Slavic University, Bishkek, Kievskaya Str.
44, 720021, Kyrgyz Republic}

\author{Vladimir Folomeev}
\email{vfolomeev@mail.ru}
\affiliation{Institute of Physics of NAS
KR, 265 a, Chui str., Bishkek, 720071,  Kyrgyz Republic}

\begin{abstract}
Three static models with two interacting phantom and ghost scalar
fields were considered: a model of a traversable wormhole, a
brane-like model and a spherically symmetric problem. It was shown
numerically that regular solutions exist for all three cases.
\end{abstract}
\maketitle

\section{Introduction}
\label{introd}
Scalar fields play a fundamental role in the modern cosmology and astrophysics. Having come in them from the theory of elementary particles, the scalar fields are widely used both at creation of models of compact objects and at research of the evolution of the Universe as a whole. In particular, the models of quasi-star objects from scalar fields, the so-called boson stars, are well-known~\cite{Ruff,Colpi,Miel}. In turn in cosmology scalar fields are the basis for creation of models of the early inflationary Universe~\cite{Linde}.

Another field of application of scalar fields is connected with one of the most exciting event in astrophysics and cosmology of the last decades - discovery of the acceleration of the present Universe~\cite{Riess, Star}. This discovery has stimulated an appearance of a large amount of various models trying to explain this phenomenon. A basis of all the models consists in violation of different energy conditions. For cosmology, most essential are two energy conditions (see, e.g., Ref.~\cite{Star}): the first one, the so-called strong energy condition (SEC), states that $\rho+3p\geq 0$, where $\rho$ and $p$ are effective energy density and pressure of matter which determine the evolution of the Universe. In hydrodynamical language this means that the parameter of the equation of state $w\geq -1/3$. This case corresponds to the decelerated expansion of the Universe (the Friedmann models). Violation of the SEC leads to the accelerated expansion of the Universe (the inflationary models). The second condition, the so-called weak energy condition (WEC), states that $\rho+p\geq 0$, whence it appears that $w\geq -1$. Failure satisfying the WEC results in exponential or more fast accelerated expansion of the Universe. The substance providing such acceleration was called dark energy. This term usually means that the equation of state lies in the range $-1 \leq w \leq -1/3$ (the cosmological constant, quintessence~\cite{Ratra}, Chaplygin gas~\cite{Kam}, the theories of
gravitation with high derivatives~\cite{Capoz}, etc.). All the models describe the present accelerated expansion of the Universe within the accuracy of observations.

The mentioned above violation of the SEC is normally arisen at some
choice of a potential energy of usual scalar fields (see examples of
such potentials in Refs.~\cite{Star}). However, as it was shown in
Ref.~\cite{Star1}  by a model-independent way on the basis of study
of data sets containing 172 SNIa, it is possible in the present
epoch that $w<-1$, i.e. the WEC can be violated.
(For more recent papers with the latest observational data see Refs.~\cite{Gong}.)
Such an unusual
state of matter is also known as phantom dark energy~\cite{Amen}.
There are models of phantom energy within the framework of
higher-order theories of gravity~\cite{Elizalde:2004mq}, braneworld
cosmology~\cite{Caldwell}, etc.

Another popular direction consists in consideration of ghost scalar
fields with negative sign before kinetic
term~\cite{Szydlowski:2004jv}. In Ref.~\cite{Dzhunush} we considered
the cosmological model of the early Universe with two interacting
 ghost scalar fields with special form of the potential
energy:
\begin{equation}
\label{pot}
V(\phi,\chi)=\frac{\lambda_1}{4}(\phi^2-m_1^2)^2+\frac{\lambda_2}{4}(\chi^2-m_2^2)^2+\phi^2
\chi^2-V_0.
\end{equation}
Here $\phi, \chi$ are two scalar fields with the masses $m_1$ and $m_2$, $\lambda_1, \lambda_2$ are the self-coupling constants and $V_0$ - some constant which should be chosen on the assumption of a problem statement. The essential feature of this potential is presence of two local minima at  $\chi~=~0, \phi~=~\pm~m_1$. It allows existence  of localized solutions with finite energy  in problems with such potential. As is known~\cite{Raj}, for a case of one scalar field, localized solutions could exist only for a case with two or more minima of potential energy when solutions start in one minimum and tend to another one. In a case of two scalar fields, a situation is possible when a solution starts and finishes  in the same minimum. In Ref.~\cite{Dzhunush} such a case was carried out: the solution started at $t\rightarrow -\infty$ and returned to the same local minimum at $t\rightarrow +\infty$. Such type of soliton-like solutions, localized on a spacelike hypersurface, known as spacelike branes (S-branes)~\cite{Sen}.

The mentioned in Ref.~\cite{Dzhunush} possibility of existence of
the localized in {\it time} solutions for phantom fields allows us
to hope for presence of similar solutions and for a {\it static}
case. In this paper we consider three models created by two
interacting phantom and ghost scalar fields with the potential
\eqref{pot}: 1) a traversable wormhole; 2) a brane-like solution --
analog of a domain wall solution but with asymptotically non-flat
spacetime (anti-de Sitter spacetime); 3) a spherically symmetric
particle-like solution.

The model of so-called traversable wormhole was suggested in
Ref.~\cite{Morris} (for a recent review, see \cite{Lobo1}). This wormhole is created by some special matter
with the violated WEC. As phantom scalar fields also violate the
WEC, they could be used at modelling of traversable wormholes. The
researches of such models were already carried out (see, e.g.,
Refs.~\cite{Sushkov, Lobo}). In these works some effective
hydrodynamical energy-momentum tensor with the equation of state
$w<-1$ was chosen as a source of matter. But a distribution of this
matter was added by hand and, correspondingly, the {\it
nonself-consistent} models of the traversable wormholes were
considered. In this paper we consider a {\it self-consistent} model
of a traversable wormhole created by two interacting phantom and
ghost scalar fields with the potential \eqref{pot}. The specified
above features of this potential allow us to find regular solutions
with localized energy density.

Domain walls are topological defects and they arise in different aspects both in particle
physics and cosmology (see, e.g., \cite{Bran2,Linde,Cvet} and references therein).
They separate a spacetime into several domains along a single coordinate. In a case of
scalar fields domain wall solutions exist when the scalar filed potentials have isolated minima.
The domain walls are surfaces separating minima of the potentials with different vacuum expectation values.
The region of fast change of the scalar field corresponds to the domain wall. The domain wall refers to as {\it thin}
wall if the energy density of the scalar field is localized
at the domain wall surface and could be replaced by the delta function.
The different variants of the domain wall solutions were found in Refs.~\cite{Berez1,Berez2,Garf,Ipser,Lagun}
with asymptotically flat, de Sitter and Schwarzschild spacetimes. There are also known the so-called {\it thick}
domain wall solutions~\cite{Arod1,Arod2,Goetz,Widr} which could exist at late-time phase transitions in
the evolution of the Universe. In this paper the consideration of the thick domain wall model with the
potential \eqref{pot} is presented.

Search of spherically symmetric solutions with various matter
sources always was an important problem in special and general
relativity. Such solutions are using both at investigation of
different particle-like models of elementary particles and creation
of models of star-like objects and another large-scale
configurations. The source of matter is fields with various spins
both interacting between each other and with gravitational field
(see, e.g., Ref.~\cite{Linde}). There are well-known regular
solutions for the scalar fields both for noninteracting and
self-interacting fields~\cite{Ruff,Colpi,Miel}. For a case of usual
(non-phantom) scalar fields, the model of boson star with potential
\eqref{pot} was considered in Ref.~\cite{Dzhun}. It was shown there
that there are regular solutions in the case under consideration.
Recently the spherically symmetric model with self-gravitating
matter  with the hydrodynamic equation of state $w<-1$ was
considered in Ref.~\cite{Shat}. Below we show that regular solutions
for spherically symmetric case exist for phantom and ghost scalar
fields also.

The paper is organized as follows: in section \ref{GE} the general gravitational and field
equations for all three above mentioned models are presented. In sections \ref{TW}, \ref{DW} and \ref{SSS}
the models of traversable wormhole, brane-like and spherically symmetric solution are considered. In section \ref{concl} we present comments and conclusions.

\section{General equations}
\label{GE}
We chose the Lagrangian as follows:
\begin{equation}
\label{lagrangian}
  L =-\frac{R}{16\pi G}+\epsilon \left[
      \frac{1}{2}\partial_\mu \varphi \partial^\mu
        \varphi + \frac{1}{2}\partial_\mu \chi \partial^\mu
        \chi-V(\varphi,\chi)
    \right]~,
\end{equation}
where $R$ is the scalar curvature, $G$ is the Newtonian gravity
constant and the constant $\epsilon =\pm 1$. In the case $\epsilon =
+1$ one has the theory of usual scalar field plus gravitation. The
case $\epsilon = -1$ corresponds to the theory of ghost scalar
field. The corresponding energy-momentum tensor will then be:
\begin{equation}
\label{emt}
    T^k_i=\epsilon \left\{
        \partial_i \varphi \partial^k \varphi+
        \partial_i \chi \partial^k \chi-
        \delta^k_i \left[
            \frac{1}{2}\partial_\mu \varphi \partial^\mu
            \varphi+\frac{1}{2}\partial_\mu \chi \partial^\mu
            \chi-V(\varphi,\chi)
        \right]
    \right\}~,
\end{equation}
and variation of the Lagrangian \eqref{lagrangian} gives the gravitational and field equations in the form:
\begin{equation}
\label{Einstein-gen}
    G_{i}^k =  8\pi G T^k_i,
\end{equation}
\begin{equation}
\label{field-gen}
\frac{1}{\sqrt{-g}}\frac{\partial}{\partial
x^\mu}\left[\sqrt{-g}\,\, g^{\mu\nu} \frac{\partial
(\varphi,\chi)}{\partial x^\nu}\right]=-\frac{\partial V}{\partial
(\varphi,\chi)}.
\end{equation}

In our case equations \eqref{Einstein-gen}-\eqref{field-gen} are the system of ordinary nonlinear differential
equations with the potential energy from \eqref{pot}. As it follows from the experience of previous
researches of similar systems~\cite{Dzhunush}, finding of solutions of the system \eqref{Einstein-gen}-\eqref{field-gen}
is reduced to search of eigenvalues of the parameters $m_1, m_2$. Procedure of search of solutions and
its application for investigation of the models mentioned in Introduction will be considered in next three
sections.

\section{Traversable wormhole}
\label{TW}

We will search for static solutions of equations \eqref{Einstein-gen}-\eqref{field-gen}
for the following metric~\cite{Visser}:
\begin{equation}
\label{metric}
ds^2=B(r) dt^2-dr^2-A(r)(d\theta^2+\sin^2\theta d\phi^2),
\end{equation}
where $A(r), B(r)$ are the even functions depending only on the coordinate $r$ which covers the entire
range $-\infty < r < +\infty$. Using this metric, one can obtain from
\eqref{Einstein-gen} and \eqref{emt} the following equations  (at
$\epsilon =-1$):
\begin{eqnarray}
\label{Einstein_a}
\frac{A^{\prime \prime}}{A}-\frac{1}{2}\left(\frac{A^{\prime}}{A}\right)^2-
\frac{1}{2}\frac{A^{\prime}}{A}\frac{B^{\prime}}{B}&=&\varphi^{\prime 2}+\chi^{\prime 2}~,  \\
\label{Einstein_b}
 \frac{A^{\prime \prime}}{A}+\frac{1}{2}\frac{A^{\prime}}{A}\frac{B^{\prime}}{B}-\frac{1}{2}\left(\frac{A^{\prime}}{A}\right)^2
 -\frac{1}{2}\left(\frac{B^{\prime}}{B}\right)^2+ \frac{B^{\prime \prime}}{B}&=&
 2\left[\frac{1}{2}(\varphi^{\prime 2}+\chi^{\prime 2})+V\right]~, \\
\label{Einstein_c}
\frac{1}{4}\left(\frac{A^{\prime}}{A}\right)^2-\frac{1}{A}+\frac{1}{2}\frac{A^{\prime}}{A}\frac{B^{\prime}}{B}&=&
 -\frac{1}{2}(\varphi^{\prime 2}+\chi^{\prime 2})+V~,
\end{eqnarray}
where a prime denotes differentiation with respect to $r$. Eq.
\eqref{Einstein_a} was obtained by subtracting $\left(^r _r\right)$ component from $\left(^t _t\right)$ component of the equations
\eqref{Einstein-gen}, and the equations \eqref{Einstein_b} and \eqref{Einstein_c} are
$\left(^\theta _\theta\right)$ and $\left(^r _r\right)$
components of Eqs. \eqref{Einstein-gen}. The corresponding field equations from \eqref{field-gen} will be:
\begin{eqnarray}
\label{field_a}
    \varphi^{\prime \prime}+\left(\frac{A^\prime}{A}+\frac{1}{2}\frac{B^\prime}{B}\right)\varphi^\prime=
    \varphi \left[2\chi^2+\lambda_1(\varphi^2-m_1^2)\right]~,  \\
    \label{field_b}
    \chi^{\prime \prime}+\left(\frac{A^\prime}{A}+\frac{1}{2}\frac{B^\prime}{B}\right)\chi^\prime=
    \chi \left[2\varphi^2+\lambda_2(\chi^2-m_2^2)\right]~.
\end{eqnarray}
In the equations \eqref{Einstein_a}-\eqref{field_b} the following rescaling are used:
$r \rightarrow \sqrt{8\pi G}\, r$, $\varphi \rightarrow \varphi/\sqrt{8\pi G}$,
$\chi \rightarrow \chi/\sqrt{8\pi G}$, $m_{1,2} \rightarrow m_{1,2}/\sqrt{8\pi G}$.

As it was shown in previous researches of problems with the potential \eqref{pot} (see, e.g.,~\cite{Dzhunush}),
regular solutions of the system of nonlinear differential equations \eqref{Einstein_a}-\eqref{field_b}
could exist only for some values of the self-coupling constants $\lambda_1, \lambda_2$ and the masses
of the scalar fields $m_1, m_2$, and also depend on boundary conditions which set under the problem statement.
Particulary, specifying some values of the parameters $\lambda_1, \lambda_2$,
one has already effect on the shape of the potential \eqref{pot} that, in turn, determine a
possibility of existence of regular solutions of the system \eqref{Einstein_a}-\eqref{field_b}.
The further task consists in a search of such parameters $m_1, m_2$ which give regular solutions.
In this sense the problem reduces to a search of {\it eigenvalues} of the parameters $m_1, m_2$  and
corresponding {\it eigenfunctions} $A, B, \varphi$ and $\chi$ for the nonlinear system of
differential equations \eqref{Einstein_a}-\eqref{field_b}.

The technique of solution of systems similar to \eqref{Einstein_a}-\eqref{field_b} is described in Ref.~\cite{dzh-step}
in details. The essence of this procedure is the following: on the first step one solve the equation \eqref{field_a}
with some arbitrary selected function $\chi$ looking for a regular solution existing only at some value of the
parameter $m_1$. On this step the influence of gravitation is not taken into account. Then this solution for the
function $\varphi$ insert into the equation \eqref{field_b} and one search for a value of the parameter
$m_2$ yielding a regular solution. This procedure reiterate several times (three usually enough) for obtaining
of acceptable convergence of values of the parameters $m_1, m_2$.
The obtained functions $\varphi, \chi$ are inserting into the gravitational equations \eqref{Einstein_a}-\eqref{Einstein_b}.
Eq. \eqref{Einstein_c}, which is the constraint equation, is using for specifying of boundary conditions
(see below). The obtained solutions for the metric functions $A, B$ are inserting into the complete equations for
the scalar fields \eqref{field_a}-\eqref{field_b} and they are solving again for a search of eigenvalues of the
parameters $m_1, m_2$ with account of gravitation. This procedure reiterate so many times as it is necessary for
obtaining of acceptable convergence of values of the parameters $m_1, m_2$.

The described procedure of a search of solutions of the system \eqref{Einstein_a}-\eqref{field_b}, also known
as the shooting method, allows to find rather fast values of the parameters $m_1, m_2$ at which regular
solutions exist. We have checked the obtained solutions using the NDSolve routine from {\it Mathematica}
substituting the eigenvalues $m_1, m_2$ and solving \eqref{Einstein_a}, \eqref{Einstein_b}, \eqref{field_a}, \eqref{field_b}
directly.

The boundary conditions are choosing with account of $\Z_2$ symmetry in the following form:
\begin{alignat}{2}
    \varphi(0)      &=\sqrt{3},               & \qquad \varphi^\prime(0)    &=0, \nonumber \\
    \chi(0)         &=\sqrt{0.6},               & \qquad \chi^\prime(0)     &=0, \nonumber \\
    A(0)         &=-\frac{1}{V(\phi(0), \chi(0))},              & \qquad A^\prime(0)        &=0, \nonumber \\
    B(0)         &=1.0,                    & \qquad B^\prime(0)        &=0,
\label{ini1}
\end{alignat}
where the condition for $A(0)$ is choosing to satisfy the constraint \eqref{Einstein_c} at $r=0$, $V(\phi(0), \chi(0))$ is the value of the potential at $r=0$ and the self-coupling constants $\lambda_1=0.1$ and $\lambda_2=1$.

\begin{figure}[t]
\begin{minipage}[t]{.49\linewidth}
  \begin{center}
  \includegraphics[width=9cm]{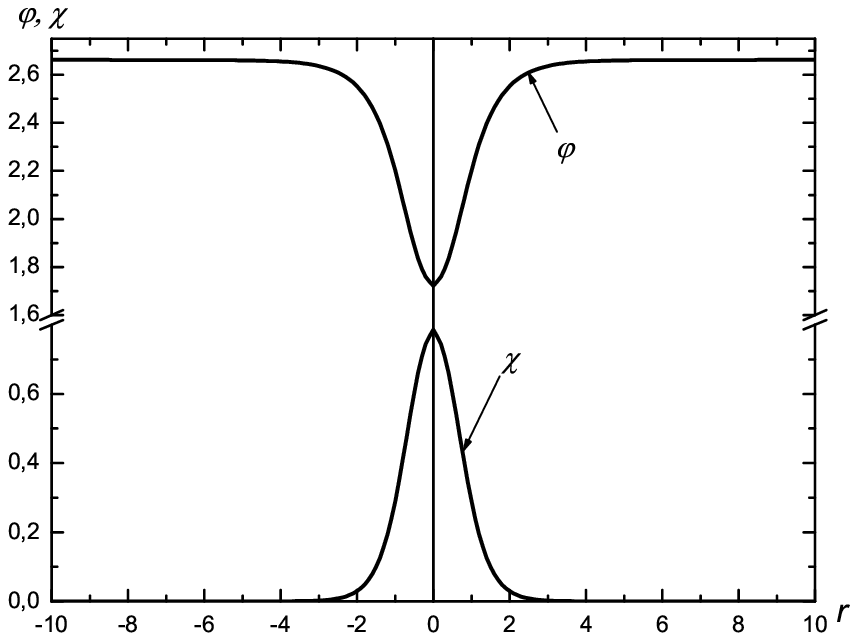}
  \caption{The scalar fields $\varphi, \chi$ in the wormhole model for the boundary conditions given in \eqref{ini1}.}
    \label{phch1}
  \end{center}
\end{minipage}\hfill
\begin{minipage}[t]{.49\linewidth}
  \begin{center}
  \includegraphics[width=9cm]{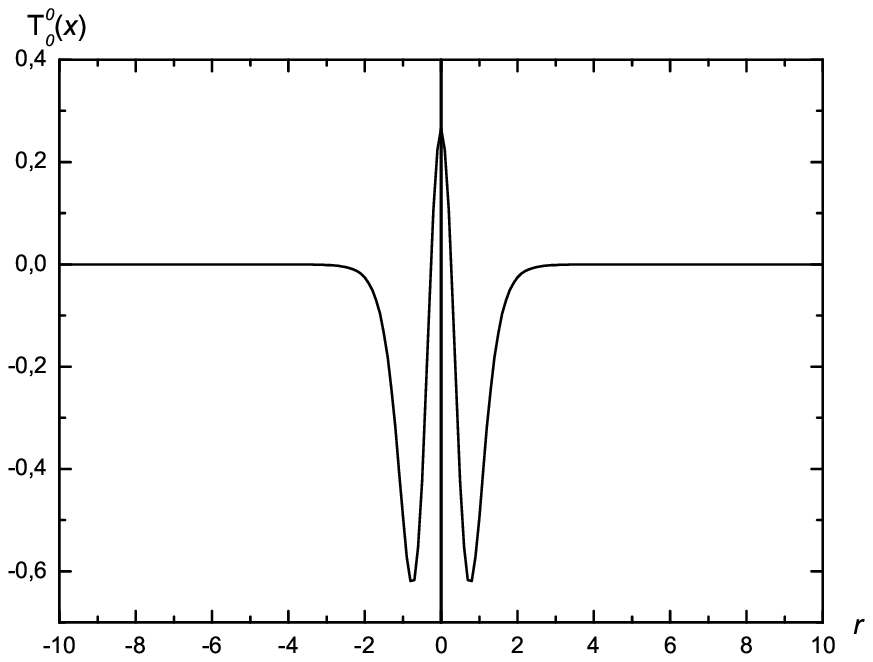}
  \caption{The energy density $T_0^0 (r)$ for the wormhole model.}
  \label{energ1}
  \end{center}
\end{minipage}\hfill
\end{figure}

\begin{figure}[t]
\begin{minipage}[t]{.49\linewidth}
  \begin{center}
  \includegraphics[width=9cm]{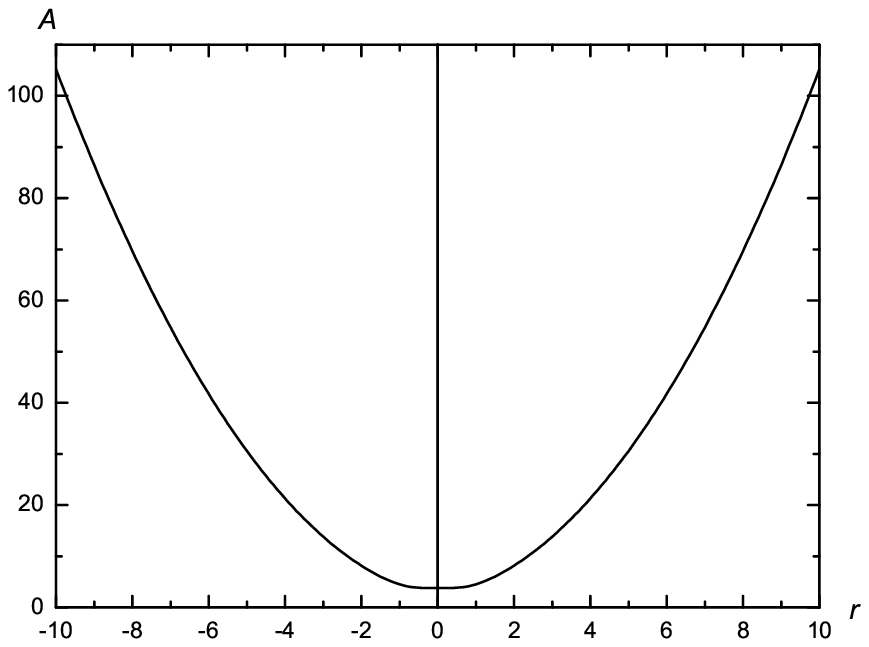}
  \caption{The metric function $A$ in the wormhole model.}
    \label{met1}
  \end{center}
\end{minipage}\hfill
\begin{minipage}[t]{.49\linewidth}
  \begin{center}
  \includegraphics[width=9cm]{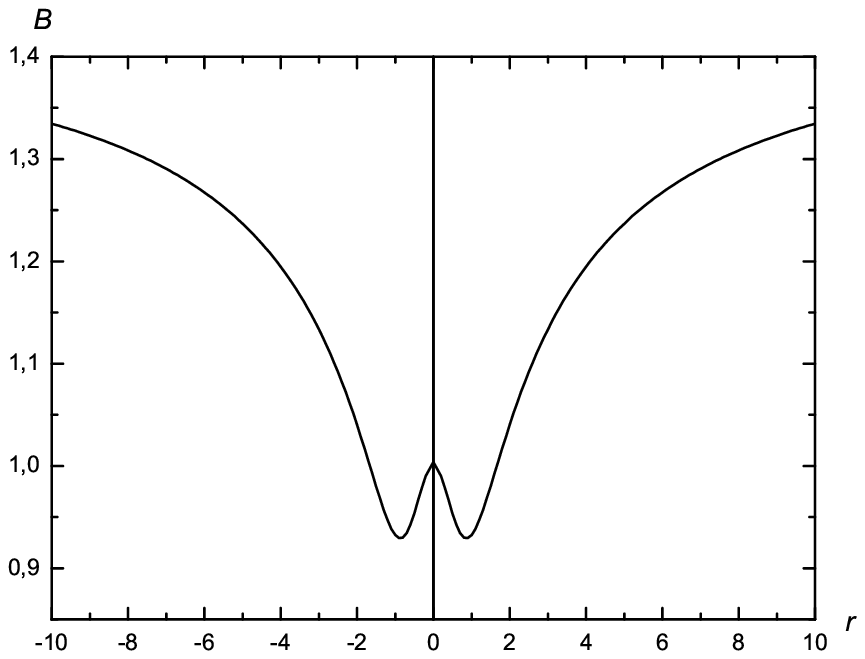}
  \caption{The metric function $B$ in the wormhole model.}
  \label{met2}
  \end{center}
\end{minipage}\hfill
\end{figure}

Then, using the above procedure for obtaining of solutions of the system \eqref{Einstein_a}-\eqref{field_b}, we
have the results presented in Fig.~\eqref{phch1}-\eqref{met2}. These results are obtained for the masses
$m_1\approx 2.661776085$ and $m_2\approx 2.928340304$. As one can see from Fig.~\eqref{phch1},
$\varphi \rightarrow m_1$ and $\chi \rightarrow 0$ as $r \rightarrow \pm\infty$. It corresponds to asymptotic transition of the solutions to the local minimum of the potential \eqref{pot} (see Introduction). The arbitrary  constant $V_0$ was chosen in such a way that the value of the potential in the local minimum was equal to zero, viz $V_0=(\lambda_2/4) m_2^4$. Such a choice of $V_0$ ensures zero value of the energy density as $r \rightarrow \pm\infty$ (Fig.~\eqref{energ1}).

Let us estimate an asymptotic behaviour of the solutions. For this purpose we will seek for solutions of the
equations \eqref{field_a}-\eqref{field_b} in the form:
\begin{equation}
\label{asymptotic}
    \varphi=m_1-\delta \varphi, \quad \chi=\delta \chi,
\end{equation}
where $\delta \varphi, \delta \chi \ll 1$ as $r \rightarrow \pm\infty$. Then the right hand side of Eqs. \eqref{Einstein_a}-\eqref{Einstein_b} goes to zero and their particular solutions are:
\begin{eqnarray}
\label{as-A}
    A &\approx&  \,r^2 + r_0^2, \\
\label{as-B}
    B &\approx& B_\infty \left( 1 - \frac{r_0^2}{r^2} \right)
\end{eqnarray}
where $r_0$ and $B_\infty$ are constants. Practically $r^2_0$ defines a total mass of a wormhole and $B_\infty$ the run of time at the infinity. By corresponding redefinition of the time $t$, these solutions could be reduced to a flat form in spherical coordinates, i.e. one has asymptotically flat Minkowski spacetime (see Figs. \eqref{met1},\eqref{met2}). Then, taken into account \eqref{as-A}-\eqref{as-B}, the corresponding asymptotic equations for the scalar fields \eqref{field_a}-\eqref{field_b} will be rewritten as:
\begin{eqnarray}
\label{as-varphi}
    \delta \varphi^{\prime \prime}+\frac{2}{r}\delta \varphi^\prime&=&2 \lambda_1 m_1^2 \delta \varphi,\\
\label{as-chi}
    \delta \chi^{\prime \prime}+\frac{2}{r}\delta \chi^\prime&=& (2 m_1^2-\lambda_2 m_2^2) \delta \chi
\end{eqnarray}
with the exponentially fast damping solutions:
\begin{eqnarray}
    \delta \varphi &\approx&  C_{\varphi} \frac{\exp{\left(- \sqrt{2 \lambda_1 m_1^2} \,\, r \right)}}{r},
\label{sol1} \\
    \delta \chi &\approx&  C_{\chi}\frac{\exp{\left(- \sqrt{ (2 m_1^2-\lambda_2 m_2^2)} \,\,r\right)}}{r},
\label{sol2}
\end{eqnarray}
where $C_{\varphi}, C_{\chi}$ are integration constants. Thus the asymptotic solutions go to vacuum ones with
the zero energy density (Fig.~\eqref{energ1}).

\section{Brane-like solution}
\label{DW}
In this section we consider a brane-like solution in 4D. Let us chose the metric in the form:
\begin{equation}
\label{metric_wall}
ds^2=a^2(x) (dt^2-dy^2-dz^2)-dx^2,
\end{equation}
where the metric function $a(x)$ depends only on the coordinate $x$.
This metric describes (2+1)-dimensional spacetime embedded in a (3+1)-dimensional spacetime.
Using \eqref{emt}, \eqref{Einstein-gen} and \eqref{field-gen},
one can obtain the $(_x^x)$ and $(_t^t)$ components of Einstein equations~\eqref{Einstein-gen}:
\begin{eqnarray}
\label{ein_wall_1}
3\left(\frac{a^\prime}{a}\right)^2&=&-\frac{1}{2}\left(\varphi^{\prime 2}+\chi^{\prime 2}\right)+V,\\
\frac{a^{\prime \prime}}{a}-\left(\frac{a^\prime}{a}\right)^2&=&\frac{1}{2}\left(\varphi^{\prime 2}+\chi^{\prime 2}\right),
\label{ein_wall_2}
\end{eqnarray}
and the scalar field equations:
\begin{eqnarray}
\label{field_wall_1}
    \varphi^{\prime \prime}+3 \frac{a^\prime}{a}\varphi^\prime&=&\varphi\left[2\chi^2+\lambda_1(\varphi^2-m_1^2)\right],\\
    \chi^{\prime \prime}+3 \frac{a^\prime}{a}\chi^\prime&=&\chi\left[2\varphi^2+\lambda_2(\chi^2-m_2^2)\right],
    \label{field_wall_2}
\end{eqnarray}
where a prime denotes differentiation with respect to $x$ and the arbitrary constant $V_0$ is choosing as follows
\begin{equation}
\label{V0_wall}
V_0=\frac{\lambda_1}{2}(\phi_0^2-m_1^2)^2+\frac{\lambda_2}{2}(\chi_0^2-m_2^2)^2+\phi_0^2 \chi_0^2
\end{equation}
for the purpose of zeroing of $a^\prime$ at $x=0$ (see Eqs.~\eqref{ein_wall_1} and \eqref{ini1_wall} below).
(Here and further we use
the same rescaling for all variables and parameters as in the previous section.)

We will solve the system of equations~\eqref{ein_wall_1}-\eqref{field_wall_2} with the
following boundary conditions at $x=0$:
\begin{alignat}{2}
\label{ini1_wall}
    \varphi(0)&=\sqrt{3},       & \qquad \varphi^\prime(0)&=0, \nonumber \\
    \chi(0)     &=\sqrt{1.8},   & \qquad \chi^\prime(0)     &=0, \\
    a(0)            &=1.0,              &  \qquad a^\prime(0)           &=0.\nonumber
\end{alignat}

\begin{figure}[ht]
\begin{minipage}[t]{.49\linewidth}
  \begin{center}
  \includegraphics[width=9cm]{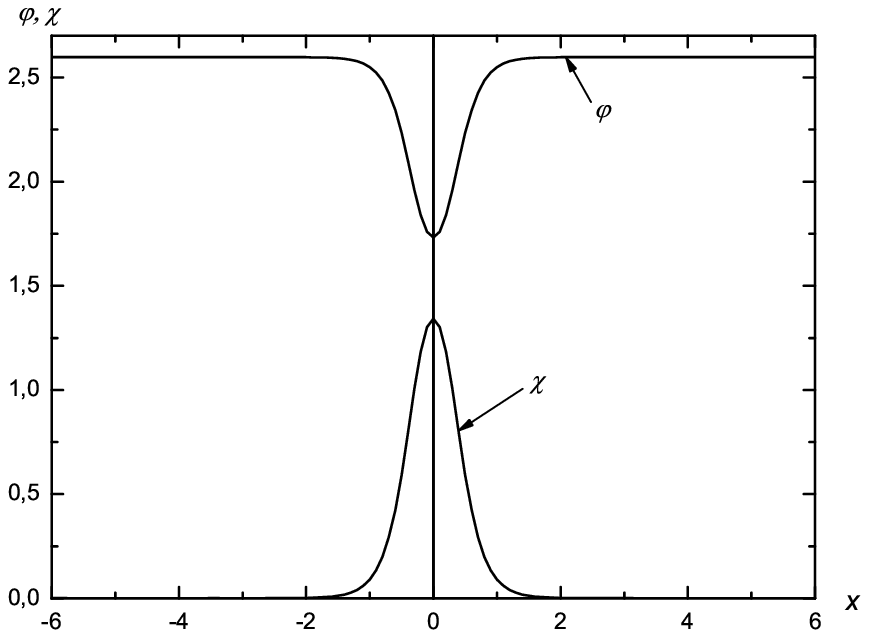}
  \caption{The distribution of the scalar fields $\phi, \chi$ near $x=0$.}
    \label{fields_wall}
  \end{center}
\end{minipage}\hfill
\begin{minipage}[t]{.49\linewidth}
  \begin{center}
  \includegraphics[width=9.5cm]{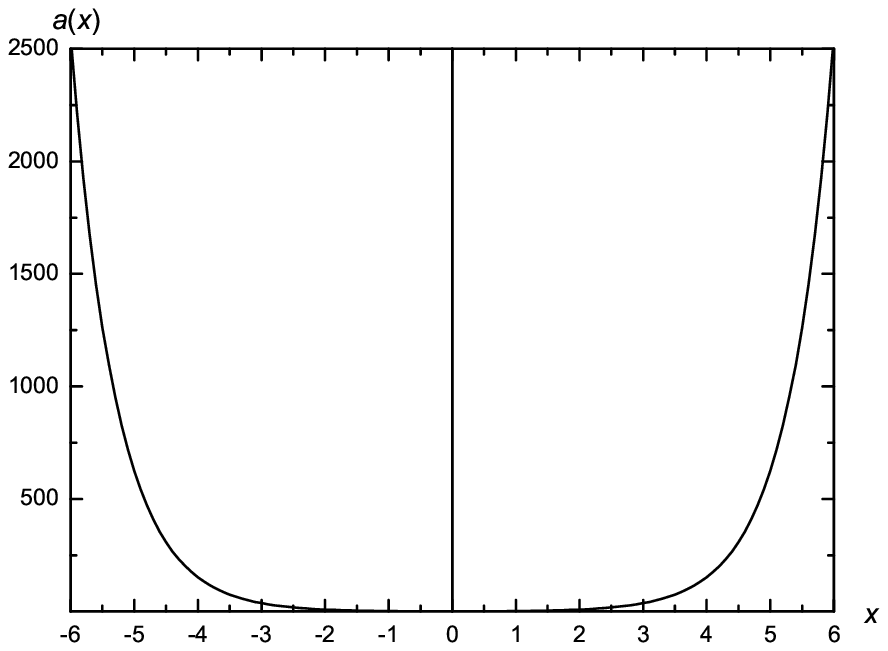}
  \caption{The metric function $a(x)$.}
  \label{mtr_wall}
  \end{center}
\end{minipage}\hfill
\end{figure}

\begin{figure}[ht]
\begin{center}
  \includegraphics[height=6cm,width=9cm]{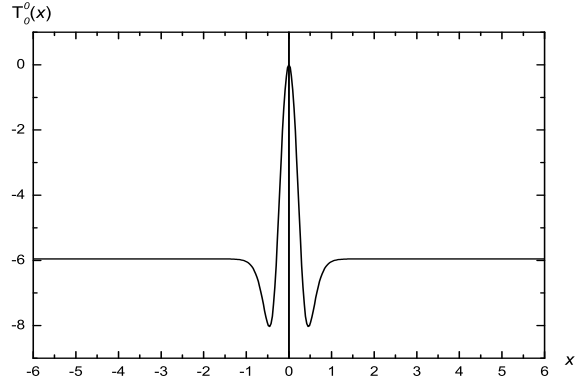}
 \caption{The localized energy density $T_0^0 (x)$ near $x=0$ with asymptotic anti-de Sitter behavior.}
\label{energy_wall}
\end{center}
\end{figure}

The procedure of finding of solutions is the same as in the previous section.
The obtained solutions with masses $m_1 \approx 2.59755, m_2 \approx 3.729$ and $\lambda_1=0.1, \lambda_2=1$
for the scalar fields are presented in Fig. \eqref{fields_wall}, for the metric function $a(x)$ in
Fig. \eqref{mtr_wall} and for the energy density $T_0^0$ in Fig. \eqref{energy_wall}.

We can easily estimate an asymptotic  behavior of the solutions. One can see from Eq.~\eqref{ein_wall_2}
that asymptotically the right hand side tends to zero and the solution of this equation is:
\begin{equation}
\label{asym_a}
a=a_0 e^{\alpha x},
\end{equation}
where $a_0$ and $\alpha$ are integration constants. This solution corresponds to the de Sitter-like solution for
the space variable $x$. Then, using \eqref{asym_a} and seeking for asymptotic solutions of the equations
\eqref{field_wall_1}-\eqref{field_wall_2} in the form:
\begin{equation}
\label{asymptotic_wall}
\varphi=m_1-\delta \varphi, \quad \chi=\delta \chi,
\end{equation}
where $\delta \varphi, \delta \chi \ll 1$ as $x \rightarrow \pm\infty$, we will have the following
equations for $\delta \varphi$ and $\delta \chi$ from~\eqref{field_wall_1}-\eqref{field_wall_2}:
\begin{eqnarray}
\label{as-varphi_wall}
\delta \varphi^{\prime \prime}+3\alpha \delta \varphi^\prime&=&2 \lambda_1 m_1^2 \delta \varphi,\\
\label{as-chi_wall}
\delta \chi^{\prime \prime}+3\alpha \delta \chi^\prime&=& (2 m_1^2-\lambda_2 m_2^2) \delta \chi
\end{eqnarray}
with the damping solutions:
\begin{eqnarray}
    \delta \varphi &\approx&  C_{\varphi} \exp{\left[-\frac{x}{2}\left(3\alpha+
    \sqrt{9\alpha^2+8\lambda_1 m_1^2}\right)\right]},
\label{sol1_wall} \\
    \delta \chi &\approx&  C_{\chi}
    \exp{\left[-\frac{x}{2}\left(3\alpha+
    \sqrt{9\alpha^2+4(2 m_1^2-\lambda_2 m_2^2)}\right)\right]},
\label{sol2_wall}
\end{eqnarray}
where $C_{\varphi}, C_{\chi}$ are integration constants. So we have the solutions that tend asymptotically to the local
minimum of the potential \eqref{pot} at $\varphi=m_1$ and $\chi=0$.

\section{Spherically symmetric solution}
\label{SSS}

For consideration of the spherically symmetric problem we take the metric in Schwarzschild coordinates:
\begin{equation}
\label{metric_sph}
ds^2=B(r)dt^2-A(r)dr^2-r^2(d\theta^2+\sin^2\theta d\phi^2).
\end{equation}
Then the $(_t^t)$ and $(_r^r)$ components of
the Einstein equations~\eqref{Einstein-gen} will be:
\begin{eqnarray}
\label{einst1_sph}
    \frac{1}{r}\frac{A^\prime}{A^2}+\frac{1}{r^2}\left(1-\frac{1}{A}\right)&=&
    -\frac{1}{2A}\left(\varphi^{\prime 2}+\chi^{\prime 2}\right)-
    V(\varphi,\chi),
\\
    \frac{1}{r}\frac{B^\prime}{A B}-\frac{1}{r^2}\left(1-\frac{1}{A}\right)&=&
    -\frac{1}{2A}\left(\varphi^{\prime 2}+\chi^{\prime 2}\right)+V(\varphi,\chi),
\label{einst2_sph}
\\
\frac{B^{\prime \prime}}{B}-\frac{1}{2}\left(\frac{B^\prime}{B}\right)^2-\frac{1}{2}\frac{A^\prime}{A}\frac{B^\prime}{B}
-\frac{1}{r}\left(\frac{A^\prime}{A}-\frac{B^\prime}{B}\right)&=& 2A\left[\frac{1}{2A}\left(\varphi^{\prime 2}+\chi^{\prime 2}\right)+
    V(\varphi,\chi)\right],
\label{einst3_sph}
\end{eqnarray}
and the scalar field equations~\eqref{field-gen} are:
\begin{eqnarray}
\label{sfe1_sph}
    \varphi^{\prime \prime}+\left(\frac{2}{r}+\frac{B^\prime}{2B}-\frac{A^\prime}{2A}\right)\varphi^\prime=
    A\varphi\left[2\chi^2+\lambda_1(\varphi^2-m_1^2)\right]~,  \\
    \chi^{\prime \prime}+\left(\frac{2}{r}+\frac{B^\prime}{2B}-\frac{A^\prime}{2A}\right)\chi^\prime=
    A\chi\left[2\varphi^2+\lambda_2(\chi^2-m_2^2)\right]~,
\label{sfe2_sph}
\end{eqnarray}
where a prime denotes differentiation with respect to $r$ and in potential~\eqref{pot}
$
V_0=\frac{\lambda_2}{4} m_2^4
$
for the purpose of zeroing of the energy density as $r\rightarrow \infty$. Equation~\eqref{einst3_sph}
is consequence of the preceding  equations~\eqref{einst1_sph}-\eqref{einst2_sph}.

\begin{figure}[t]
\begin{minipage}[t]{.49\linewidth}
  \begin{center}
  \includegraphics[width=9cm]{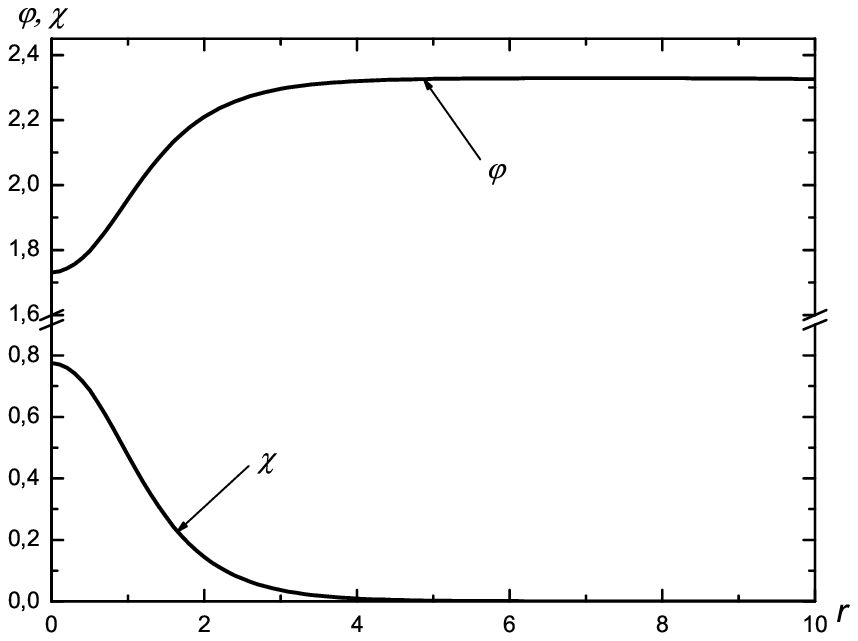}
  \caption{The scalar fields $\phi, \chi$ for the spherically symmetric case.}
    \label{fields_sph}
  \end{center}
\end{minipage}\hfill
\begin{minipage}[t]{.49\linewidth}
  \begin{center}
  \includegraphics[width=9cm]{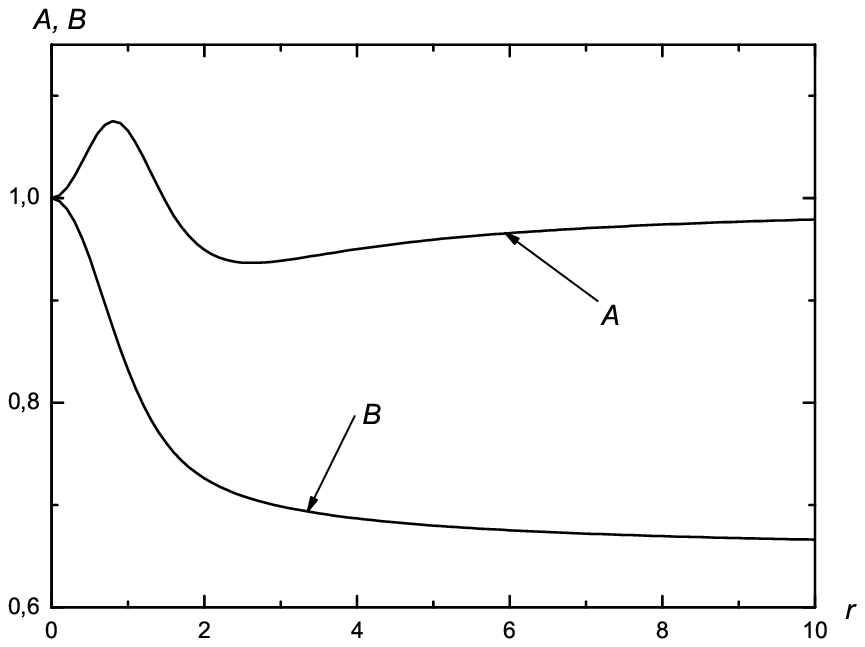}
  \caption{The metric function $A(r), B(r)$ for the spherically symmetric case.}
  \label{mtr_sph}
  \end{center}
\end{minipage}\hfill
\end{figure}

\begin{figure}[t]
\begin{minipage}[t]{.49\linewidth}
  \begin{center}
  \includegraphics[width=9cm]{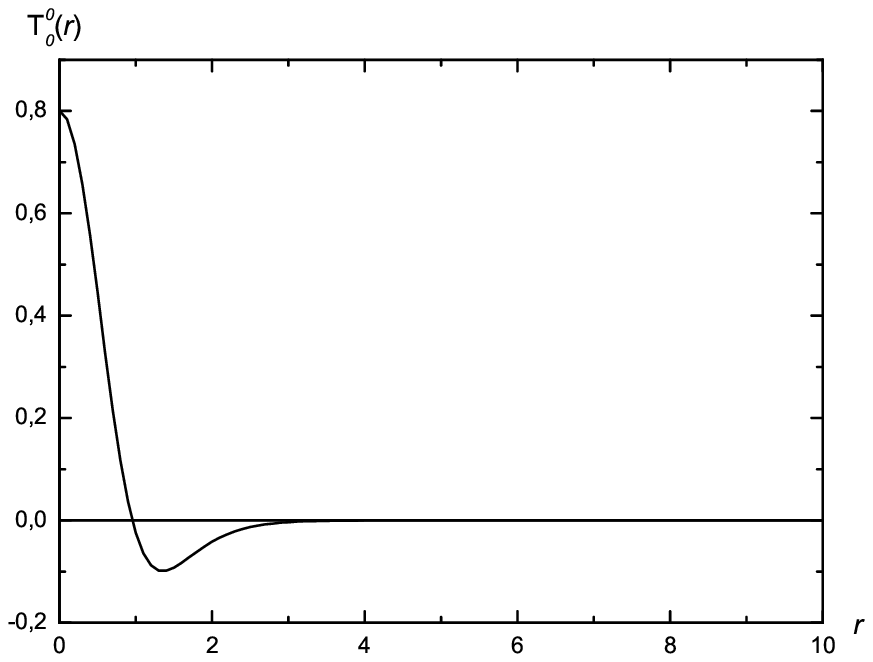}
  \caption{The energy density $T_0^0 (r)$ for the spherically symmetric case.}
    \label{energy_sph}
  \end{center}
\end{minipage}\hfill
\begin{minipage}[t]{.49\linewidth}
  \begin{center}
  \includegraphics[width=9cm]{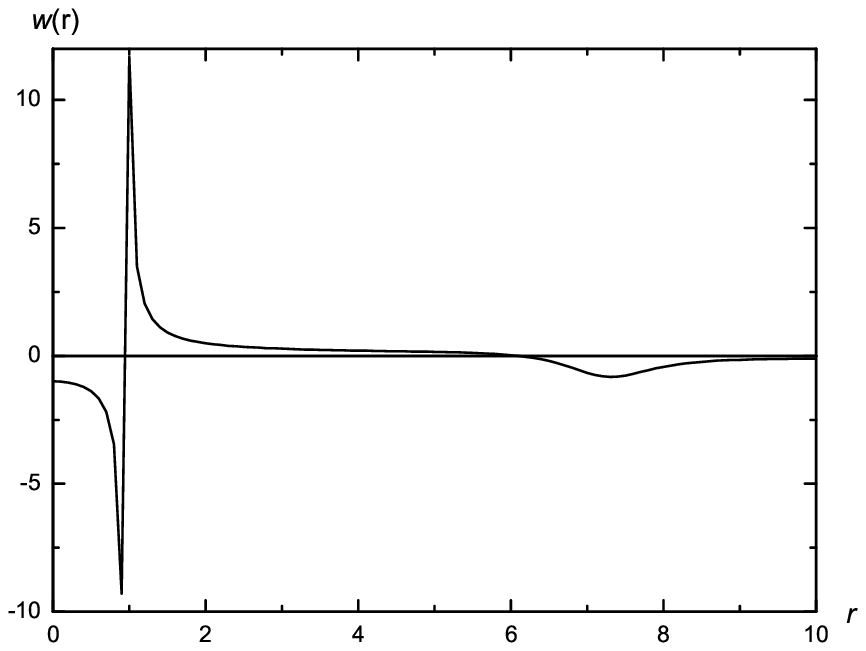}
  \caption{The equation of state $w(r)$ for the spherically symmetric case.}
  \label{eqs_f}
  \end{center}
\end{minipage}\hfill
\end{figure}

Choosing the boundary conditions at $r=0$ in the form:
\begin{alignat}{2}
    \label{ini1_sph}
    \varphi(0)&=\sqrt{3},& \qquad \varphi^\prime(0)&=0, \nonumber \\
    \chi(0)&=\sqrt{0.6},& \qquad \chi^\prime(0)&=0, \\
    A(0)&=1.0,&  \qquad B(0)&=1.0\nonumber
\end{alignat}
and following the above procedure of obtaining of solutions
we find the masses $m_1 \approx 2.329305, m_2 \approx 3.0758999$ at
$\lambda_1=0.1, \lambda_2=1$. The results of numerical calculations
for the scalar fields are presented in Fig. \eqref{fields_sph}, for the metric functions $A(r), B(r)$ in
Fig. \eqref{mtr_sph} and for the energy density $T_0^0$ in Fig. \eqref{energy_sph}.

One can see from \eqref{einst1_sph}-\eqref{einst2_sph} that asymptotic behavior of the metric functions $A(r)$ and $B(r)$ is:
\begin{equation}
\label{asym_sph}
    A \approx \frac{1}{1 + \frac{r_0}{r}}, \quad
    B \approx B_\infty \left( 1 + \frac{r_0}{r} \right)
\end{equation}
where $r_0$ and $B_\infty$ are constants. Practically $r_0$ defines a total mass and $B_\infty$ the run of time at the infinity. Redefining the time variable $t$ we can chose $B=1$ as $r\rightarrow \infty$. I.e. we have asymptotically flat Minkowski spacetime. The corresponding asymptotic scalar field  equations~\eqref{sfe1_sph}-\eqref{sfe2_sph} with account of
\begin{equation}
\label{asymptotic_sss}
\varphi=m_1-\delta \varphi, \quad \chi=\delta \chi
\end{equation}
will be:
\begin{eqnarray}
\label{as-varphi_sss}
\delta \varphi^{\prime \prime}+\frac{2}{r}\delta \varphi^\prime&=&2 \lambda_1 m_1^2 \delta \varphi,\\
\label{as-chi_sss}
\delta \chi^{\prime \prime}+\frac{2}{r}\delta \chi^\prime&=& (2 m_1^2-\lambda_2 m_2^2) \delta \chi
\end{eqnarray}
with the exponentially fast damping solutions:
\begin{eqnarray}
\delta \varphi &\approx&  C_{\varphi} \frac{\exp{\left(- \sqrt{2 \lambda_1 m_1^2} \,\, r \right)}}{r},
\label{sol1b} \\
\delta \chi &\approx&  C_{\chi}\frac{\exp{\left(- \sqrt{ (2 m_1^2-\lambda_2 m_2^2)} \,\,r\right)}}{r},
\label{sol2b}
\end{eqnarray}
where $C_{\varphi}, C_{\chi}$ are integration constants. Thus the asymptotic solutions go to vacuum ones with
the zero energy density (Fig.~\eqref{energy_sph}).

Finally, let us show evolution of the effective equation of state $w(r)=p(r)/\varepsilon(r)$ where $\varepsilon(r)$ and
$p$ are the effective energy density and pressure of the scalar fields. In the case under consideration we have
from~\eqref{emt} and \eqref{metric_sph}:
\begin{eqnarray}
T^0_0&=&\varepsilon(r)=-\left[\frac{1}{2A}\left(\varphi^{\prime 2}+\chi^{\prime 2}\right)+V(\varphi,\chi)\right],\\
T^1_1&=&-p(r)=-\left[-\frac{1}{2A}\left(\varphi^{\prime 2}+\chi^{\prime 2}\right)+V(\varphi,\chi)\right].
\end{eqnarray}
Then the corresponding effective equation of state will be:
\begin{equation}
\label{eqs}
w(r)=p(r)/\varepsilon(r)=-\frac{-\frac{1}{2A}\left(\varphi^{\prime 2}+\chi^{\prime 2}\right)+V(\varphi,\chi)}
{\frac{1}{2A}\left(\varphi^{\prime 2}+\chi^{\prime 2}\right)+V(\varphi,\chi)}.
\end{equation}
Using the numerical solution obtained earlier, we have the following graph for the equation of state (see Fig.~\eqref{eqs_f}).
As one can see from this figure, there is some point $r=r_*$ in which the denominator in~\eqref{eqs} tends to zero
(in the case under consideration $r_*\approx 0.956$).
In the range $0<r<r_*$ we have $w\leq -1$ ($w=-1$ at $r=0$), and in the range $r_*<r<10$ we have $w>-1$. At $r\rightarrow r_*$ from
the left $w\rightarrow -\infty$ and at $r\rightarrow r_*$ from
the right $w\rightarrow +\infty$.
Asymptotically $w(r)$ tends to zero, i.e. we have the dust-like equation of state.

It follows from this that in the entire range $0<r<r_*$ the WEC is violated, i.e. efficiently the scalar fields describe
phantom matter. On the other hand in the range $r>r_*$ the WEC is not violated and the scalar fields are non-phantom.
The scalar fields with the equation of state varying in time are used in the theories describing the present acceleration
of the Universe (see, e.g., \cite{Kamen}).
In such theories there is some point of time in which the violation of the WEC and transition of
usual scalar fields into phantom ones take place. In our case we deal with inhomogeneous distribution of the scalar fields with
the varying in space equation of state and transition of
usual scalar fields into phantom ones in some point $r=r_*$.

\section{Conclusion}
\label{concl} We have considered three static solutions for the
model of two interacting phantom and ghost scalar fields: the model
of a traversable wormhole, the brane-like model and the spherically
symmetric problem. The self-consistent problems with account of the
back reaction of the scalar fields on gravitation were investigated.
The choice of a potential in the form \eqref{pot} ensures existence
of two local minima that allow to find regular solutions which start
and finish in one of these minima. In this case the nonlinear
problems on evaluation of eigenvalues of the parameters $m_1, m_2$
ensuring the mentioned regular solutions were solved. Note that
existence of such solutions depends on values of the self-coupling
constants $\lambda_1, \lambda_2$ and the boundary conditions. There
is some range of these parameters appropriate for existence of
regular solutions. Particularly, this range is defining by
conditions on existence of the local and global minima of the
potential \eqref{pot}: $\lambda_1>0, m_1^2>\lambda_2 m_2^2/2$;
$\lambda_2>0, m_2^2>\lambda_1 m_1^2/2$; $\lambda_2 m_2^4> \lambda_1
m_1^4$.

For the wormhole model, it was shown that there exist regular solutions in the entire range $-\infty < r < +\infty$ with asymptotically flat Minkowski spacetime (see Eqs. \eqref{as-A}, \eqref{as-B}). I.e. the obtained wormhole solution connects two flat spacetimes. The radius of the wormhole throat is defined by a minimal  value of the function $A(r)$, i.e.
$R_0=\min \limits_{r \in (-\infty, +\infty)} \{A(r)\}$ at $r=0$. As one can see in Fig.~\eqref{energ1},
for a remote observer the phantom energy looks like a compact object localized near the throat of the wormhole with some negative energy density.

For the brane-like case, the obtained solutions describe (2+1)-dimensional spacetime embedded in a
(3+1)-dimensional spacetime. The asymptotic value of the potential~\eqref{pot}, $V(x=\pm \infty)<0$ plays the
role of a negative cosmological constant, and so the metric~\eqref{metric_wall} is asymptotically anti-de Sitter with
corresponding anti-de Sitter horizon. In this sense such a solution differs from domain wall solution which has flat asymptotics.

For the spherically symmetric case we have found particle-like
solutions with asymptotically flat Minkowski spacetime. It was shown
that the effective equation of state $w(r)$ for the scalar fields
changes essentially along the radius $r$ (see Fig.~\eqref{eqs_f}).
There exists some point $r=r_*$ dividing the whole space into two
regions: in the entire range $0<r<r_*$, $w<-1$ that corresponds to
phantom-like behavior of the scalar fields. On the other hand, in
the range $r_*<r<\infty$ the equation of state $w>-1$ that
corresponds to non-phantom (ghost) scalar fields. Similar behavior
of the equation of state could be obtained both for the wormhole and
brane-like models as well.

Let us note that for all three models, the obtained solutions correspond to soliton-like solutions starting and finishing in the same minimum (in the cases under consideration, in the local minimum $\phi=m_1, \chi=0$ of the potential~\eqref{pot}). In models with one scalar field soliton-like solutions exist only in a case of presence of two or more minima and they start from one minimum of a potential and tend asymptotically to another one. By the
terminology of~\cite{Raj}, such solutions refer to topological solutions, and our solutions - to non-topological ones.

Note here that the known constraint on possibility of existence of regular static solutions in spaces with
dimensionality $D\geq 3$ does not work in our case. As it was shown in Ref.~\cite{derrick} (see also Ref.~\cite{Raj}),
the mentioned solutions for usual (nonphantom) scalar fields do not exist if the potential $V\geq 0$ in the whole space.
On the other hand, for phantom fields the condition $V\leq 0$ should be satisfied always (see in this connection Ref.~\cite{Bron}). In our case the potential \eqref{pot} changes its sign that allow to avoid the constraint of the theorem from Ref.~\cite{derrick} and obtain static regular solutions.

Our attempts to find regular solutions with $\epsilon=+1$ were not succeeded. However, this does not exclude a
possibility of their existence at some parameters $\lambda_1, \lambda_2$ in the potential \eqref{pot} and some
special boundary conditions.

\section*{Acknowledgments}
The authors would like to thank the referee for helpful comments and for some suggested references.

\end{document}